\documentclass[prb,fleqn,12pt,showpacs]{revtex4-1}
\usepackage{epsfig}
\usepackage{graphicx}
\usepackage{amsfonts,amssymb}
\usepackage{amsmath}

\newcommand{\be}{\begin{equation}}
\newcommand{\ee}{\end{equation}}   
\newcommand{\bea}{\begin{eqnarray}}
\newcommand{\eea}{\end{eqnarray}}
\newcommand{\ba}{\begin{array}}
\newcommand{\ea}{\end{array}}
\newcommand{\phr}[1]{Phys.~Rev. {\bf #1}}
\newcommand{\phrl}[1]{Phys.~Rev.~Lett. {\bf #1}}
\newcommand{\phrb}[1]{Phys.~Rev.~B {\bf #1}}

\newcommand{\q}{{\bf q}}

\begin{document}
\title{Orbital ordering transition in the single-layer manganites near half doping: a weak-coupling approach}
\author{Dheeraj Kumar Singh$^1$}
\author{Tetsuya Takimoto$^2$}
\email{takimoto@hanyang.ac.kr}
 
\affiliation{$^1$Asia Pacific Center for Theoretical Physics, Pohang, Gyeongbuk 790-784, Korea}
\affiliation{$^2$Department of Physics, Hanyang University, 
17 Haengdang, Seongdong, Seoul 133-791, Korea}

\begin{abstract}
The roles of crystal-field splitting and Jahn-Teller distortions on the orbital ordering transition are investigated in the single-layer manganites near half doping. Crystal-field splitting of energy levels favoring the $d_{3z^2-r^2}$ occupancy provides not only the
 correct Fermi surface topology for  La$_{0.5}$Sr$_{1.5}$MnO$_4$ having
 a circular electron pocket around the $\Gamma$ point, but also enhances the flatness of the hole pocket around the M point thereby improving 
the nesting. In the presence of the circular electron pocket, Jahn-Teller distortions are found
 to be crucial for the transition to the transverse orbital ordering with ordering wavevector ($0.5\pi, 0.5\pi$). In the hole-doping regime
 $0.5 \leq x \leq 0.7$, the orbital ordering wavevector shows a linear dependence on the hole concentration 
in accordance with the experiments. 
\end{abstract}
\maketitle
\newpage
\section{Introduction}
The intricate interplay of spin, charge, orbital, and lattice degrees of freedom is highlighted exquisitely by
 CE-type phase, which exhibits simultaneous spin, charge, and orbital ordering in half-doped layered\cite{moritomo,zheng} as well as pseudo-cubic manganites\cite{kajimoto}. 
The composite spin-charge-orbital ordered state consists of Mn$^{3+}$ and Mn$^{4+}$ ions arranged in a checkerboard pattern,
 an orbital order of so-called $e_g$ electrons of the Mn$^{3+}$ ions with wave vector $(\pi/2,\pi/2,0)$, and a ferromagnetic alignment of spins of Mn$^{3+(4+)}$ ions along a zig-zag
 chain with the antiferromagnetic coupling between the neighboring chains.\cite{wollan,goodenough} 

Half-doped La$_{0.5}$Sr$_{1.5}$MnO$_4$ exhibits a charge and orbital ordering at a transition 
temperature $T_{OO}$ $\approx$ 220 K, and undergoes a further phase transition to the CE-type ordered state at a lower temperature $T_{N}$ $\approx$ 110 K retaining
 the charge and orbital structures found below $T_{OO}$. The charge-orbital ordering observed in the hole-doped regime 0.5 $\leq$ $x$ $<$ 0.7 for the single-layer manganites by
 high-resolution electron microscopy (HREM)\cite{nagai}, optical spectroscopy \cite{lee}, and x-ray experiments \cite{Larochelle}  
has been described as 'Wigner crystal' type\cite{radaelli} associated with a charge-density wave of $d_{x^2-y^2} (d_{3z^2-r^2})$ electrons. The orbital ordering wavevector depends
 on the hole doping $x$ in an elementary manner $Q_x = Q_y = \pi(1-x)$.   

Nature of the orbital ordering have been explored using resonant elastic soft x-ray scattering experiments (RSXS) \cite{murakami,dhesi,wilkins} and 
linear dichroism (LD)\cite{huang,wu} in La$_{0.5}$Sr$_{1.5}$MnO$_4$. However, no consensus has been reached yet
 regarding whether $d_{3x^2-r^2}$/$d_{3y^2-r^2}$\cite{dhesi,wu}- or $d_{x^2-z^2}$/$d_{y^2-z^2}$\cite{huang,wilkins}-type
 orbital order exists. On the other hand, both the experimental methods have emphasized the important role of Jahn-Tellar distortions of MnO$_6$ octahedra on stabilizing the orbital order.\cite{huang, wilkins}  

In the undoped LaSrMnO$_4$, two-fold degeneracy of $d_{x^2-y^2} (d_{3z^2-r^2})$ orbitals is no longer present in the tetragonal symmetry, which results in a favorable occupancy of $d_{3z^2-r^2}$ orbital in 
comparison to $d_{x^2-r^2}$ as revealed by x-ray diffraction (XRD) \cite{merz} and optical spectra measurements.\cite{moritomo_jpsj} The estimate of the effective crystal-field parameter $\Delta$ by the optical spectra
 measurement is around $\sim$ 0.5 eV, which is of the order of hopping parameter according to the band-structure calculation.\cite{park} However, the difference in the occupancies of in- and 
out-of-plane orbitals diminishes on increasing Sr doping until it becomes small at half doping. This can result from an additional contribution of the long-range Coulomb interaction to the crystal field, wherein nearest neighbor Mn$^{4+}$ ions tend to drag the out-of-plane electron
 cloud to the basal plane, thereby enhancing the occupancy of $d_{x^2-y^2}$ orbital.\cite{sboychakov} Therefore, the effective crystal-field parameter 
 at half doping must be smaller than the undoped case. 

Most of the theoretical studies investigating the charge and orbital ordering in manganites have been carried out in the strong coupling regime.\cite{kugel} However, 
recent angle resolved photoemission spectroscopy (ARPES) measurements\cite{chuang,mannela,evtushinsky} on layered manganites have suggested an alternate plausible 
explanation for the charge-orbital ordered state discussed above, wherein the relationship between the Fermi surface characteristics and the charge/orbital order is crucial. According to the ARPES measurement, the Fermi surface of La$_{0.5}$Sr$_{1.5}$MnO$_4$ consists of a circular electron pocket around the $\Gamma$ point and a relatively large hole pocket around 
the M point.\cite{evtushinsky} The flatness of substantial portion of the hole pockets provides good nesting, which can lead eventually to the Fermi surface instability
 to an orbital order with renormalized interactions within the Fermi-liquid picture. Furthermore, the nesting wavevector shows a linear dependence on the hole doping 
and is incommensurate within the hole-doping regime
 0.4 $\leq$ $x$ $<$ 0.6 in the bilayer manganites\cite{zsun} whereas similar ARPES measurements are unavailable for the single-layer manganites. In a recent study of the density-wave state proposed 
for the half-doped single layer manganite, only the roles of on-site inter-orbital and inter-site Coulomb interaction have been investigated while the electron 
pocket around the $\Gamma$ point observed by ARPES was not considered.\cite{yao} 

In this paper, we emphasize on the essential role of Jahn-Teller distortions in obtaining the experimentally observed orbital ordering in the charge-orbital
 ordered phase near half doping for the realistic electronic state observed by ARPES, which also includes the electron pocket around the $\Gamma$ point. To reproduce ARPES Fermi surface, we incorporate the tetragonal crystal-field splitting between $d_{x^2-y^2}$ and $d_{3z^2-r^2}$ orbitals, which can significantly 
influence the shape as well as the orbital composition of Fermi surface, and hence the orbital ordering. We focus only on the orbital aspect of the transition 
involving ordering parameters belonging to either B$_{1g}$ representation which breaks the four-fold rotation symmetry or A$_{1g}$ 
representation which does not, without considering the charge
 ordering which can be induced in the orbital ordered state by the long-range Coulomb interaction. 
Moreover, the dependence of orbital instability on the hole doping is investigated in the entire hole doping region 0.5 $\leq$ $x$ $\leq$ 0.7.
\section{Model Hamiltonian}
We consider a two orbital Hubbard-type Hamiltonian spanned by $d_{x^2-y^2}$ and $d_{3z^2-r^2}$ orbitals for the single-layer manganites
\begin{eqnarray}
{\mathcal H} &=&H_{\rm kin}+H_{\rm el-el}+H_{\rm CEF}+H_{\rm JT},
\end{eqnarray}
which includes kinetic term $H_{\rm kin}$, on-site Coulomb interaction $H_{\rm el-el}$, tetragonal crystal-field splitting $H_{\rm CEF}$, and Jahn-Teller term $H_{\rm JT}$.

The kinetic term within the tight-binding description
\begin{eqnarray}
  H_{\rm kin} &=& -\sum_{\gamma \gamma' \sigma  {\bf ia}}
  t^{\bf a}_{\gamma \gamma'} d_{ \gamma \sigma {\bf i}}^{\dag}
  d_{\gamma' \sigma  {\bf i+a}} 
\end{eqnarray}
includes $d^{\dagger}_{1 \sigma {\bf i}}$ ($d^{\dagger}_{2 \sigma {\bf i}}$) as the electron creation operator at site ${\bf i}$ with spin $\sigma$ in the orbital $d_{x^2-y^2}$ ($d_{3z^2-r^2}$). 
$t^{\bf a}_{\gamma \gamma'}$ are the
 hopping elements between $\gamma$ and $\gamma'$ orbitals along ${\bf a}$ 
connecting the nearest-neighboring sites, 
which are given by
$t^{\bf x}_{11}$ = $-\sqrt{3}t^{\bf x}_{12}$ = $-\sqrt{3}t^{\bf x}_{21}$ = $3t^{\bf x}_{22}$ = $3t/4$ 
for ${\bf a}$ = ${\bf x}$ and
$t^{\bf y}_{11}$ = $\sqrt{3}t^{\bf y}_{12}$ = $\sqrt{3}t^{\bf y}_{21}$ = $3t^{\bf y}_{22}$ = $3t/4$
for ${\bf a}$ = ${\bf y}$, respectively. In the following, $t$ is set to be the unit of energy.

The 
crystalline-electric field (CEF) 
term accounts for the splitting of $e_g$ levels in the tetragonal symmetry, and is given by
\begin{equation}
H_{\rm CEF}= -\Delta \sum_{\bf i}\mathcal{T}^z_{\bf i} = -\Delta \sum_{\sigma  {\bf i}} (d_{1 \sigma {\bf i}}^{\dag} d_{ 1 \sigma  {\bf i}}-d_{2 \sigma  {\bf i}}^{\dag} d_{2 \sigma {\bf i}}).
\end{equation}
A negative $\Delta$ which favors the occupancy of $d_{3z^2-r^2}$ orbital over $d_{x^2-y^2}$ orbital is used herefrom.

The on-site Coulomb interaction 
\begin{eqnarray}
 H_{\rm el-el} = U \sum_{\gamma {\bf i}} n_{ \gamma  \sigma {\bf i}}
  n_{\gamma -\sigma  {\bf i}} +U'\sum_{{\bf i}} n_{\gamma {\bf i} }n_{\gamma' {\bf i}}
-J_{\rm H} \sum_{{\bf i}} {\bf S}_{{\bf i}} \cdot {\bf s}_{{\bf i}}
\end{eqnarray}
includes intra-orbital ($U$) and inter-orbital ($U'$) Coulomb interactions. Third term represents the 
Hund's coupling ($J_{\rm H}$) between the spin ${\bf s}_i=\sum_{\gamma \sigma \sigma'} d_{  \gamma \sigma {\bf i}}^{\dag} {\pmb{\sigma}}_{\sigma \sigma'} d_{  \gamma \sigma' {\bf i}}$ of $e_g$ electrons and the
 localized $t_{2g}$ spin ${\bf S}_i$. Since the spins are thermally disordered in the high-temperature phase of charge-orbitally ordered state, we drop the third term for simplicity 
while keeping the intra-orbital Coulomb interaction term which, apart from being the largest interaction in the manganites, can also influence the orbital ordering as discussed below. 

Finally, we consider the Jahn Teller term 
\begin{eqnarray}
  H_{\rm JT} &=& \sum_{l {\bf i}}
 gQ_{l{\bf i}} \mathcal{T}^l_{{\bf i}}+ \sum_{l \bf i}[ P^2_{l {\bf i}}/(2M)+K_l Q^2_{l {\bf i}}/2],
\end{eqnarray}
wherein $Q_{0 {\bf i}}$, $Q_{1 {\bf i}}$ and $Q_{2 {\bf i}}$ are the breathing mode distortion,
($x^2-y^2$)- and ($3z^2-r^2$)-type Jahn-Teller distortions, respectively. $P_{l {\bf i}}$ is the canonical cojugate momentum of $Q_{l {\bf i}}$.  Here,
\begin{eqnarray}
\mathcal{T}^0_{{\bf i}} &=& \sum_{  \gamma \sigma {\bf i}} d_{ \gamma \sigma {\bf i} }^{\dag} d_{ \gamma \sigma {\bf i}} \nonumber\\
\mathcal{T}^x_{{\bf i}} &=& \sum_{{\bf i} \sigma} (d_{1 \sigma  {\bf i}}^{\dag} d_{2 \sigma  {\bf i}}+d_{2 \sigma  {\bf i}}^{\dag} d_{1 \sigma {\bf i}}) \nonumber\\ 
\mathcal{T}^z_{{\bf i}} &=& \sum_{{\bf i} \sigma} (d_{1 \sigma  {\bf i}}^{\dag} d_{1 \sigma  {\bf i}}-d_{2 \sigma {\bf i}}^{\dag} d_{2 \sigma {\bf i}}) 
\end{eqnarray}
are the charge operator, the transverse and longitudinal components of orbital operator, respectively.
The second term of $H_{JT}$ represents the kinetic and the potential energies for the distortions with the common mass $M$ and the spring constant $K_l$. In the following, we assume $K_x = K_z = fK_0 = K$ though our system has the tetragonal symmetry, where $f$ is the ratio of
spring constants for breathing and Jahn-Teller modes. Using the standard quantization of phonons with setting $\hbar = 1$, the phonon operator $a_{l{\bf i}}$ is related to the distortion $Q_{l {\bf i}}$ through $Q_{l {\bf i}}$ = 
$(a_{l {\bf i}}$+$a^{\dag}_{l {\bf i}}$)/$\sqrt{2 \omega^{\prime}_l M}$ with the phonon energy 
$\omega^{\prime}_l = \sqrt{K_l/M}$, so that the Jahn-Teller term can be expressed as   
\begin{eqnarray}
   H_{\rm JT} &=& \sum_{l {\bf i}} g^{\prime}_l (a_{l {\bf i}}+a^{\dag}_{l{\bf i}})\mathcal{T}^l_{ {\bf i}}+\sum_{l {\bf i}} \omega^{\prime}_l (a^{\dag}_{l {\bf i}}a_{l {\bf i}}+1/2),
\end{eqnarray}
where $g^{\prime}_l = g/\sqrt{2 \omega^{\prime}_l M}$.
\section{Electronic States}
\begin{figure}
\begin{center}
\vspace*{-2mm}
\hspace*{-0mm}
\psfig{figure=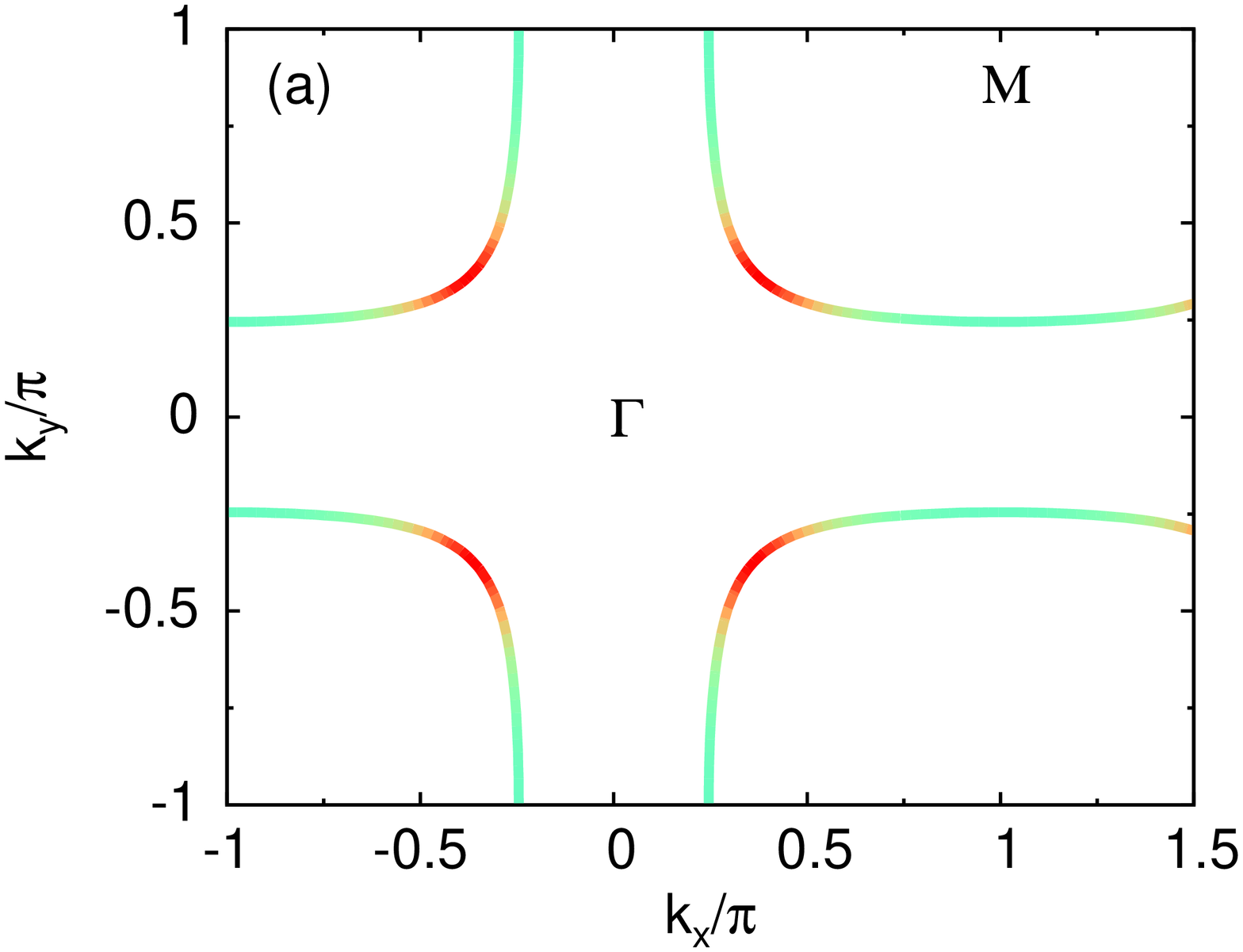,width=55mm,angle=-90}
\hspace*{5mm}
\psfig{figure=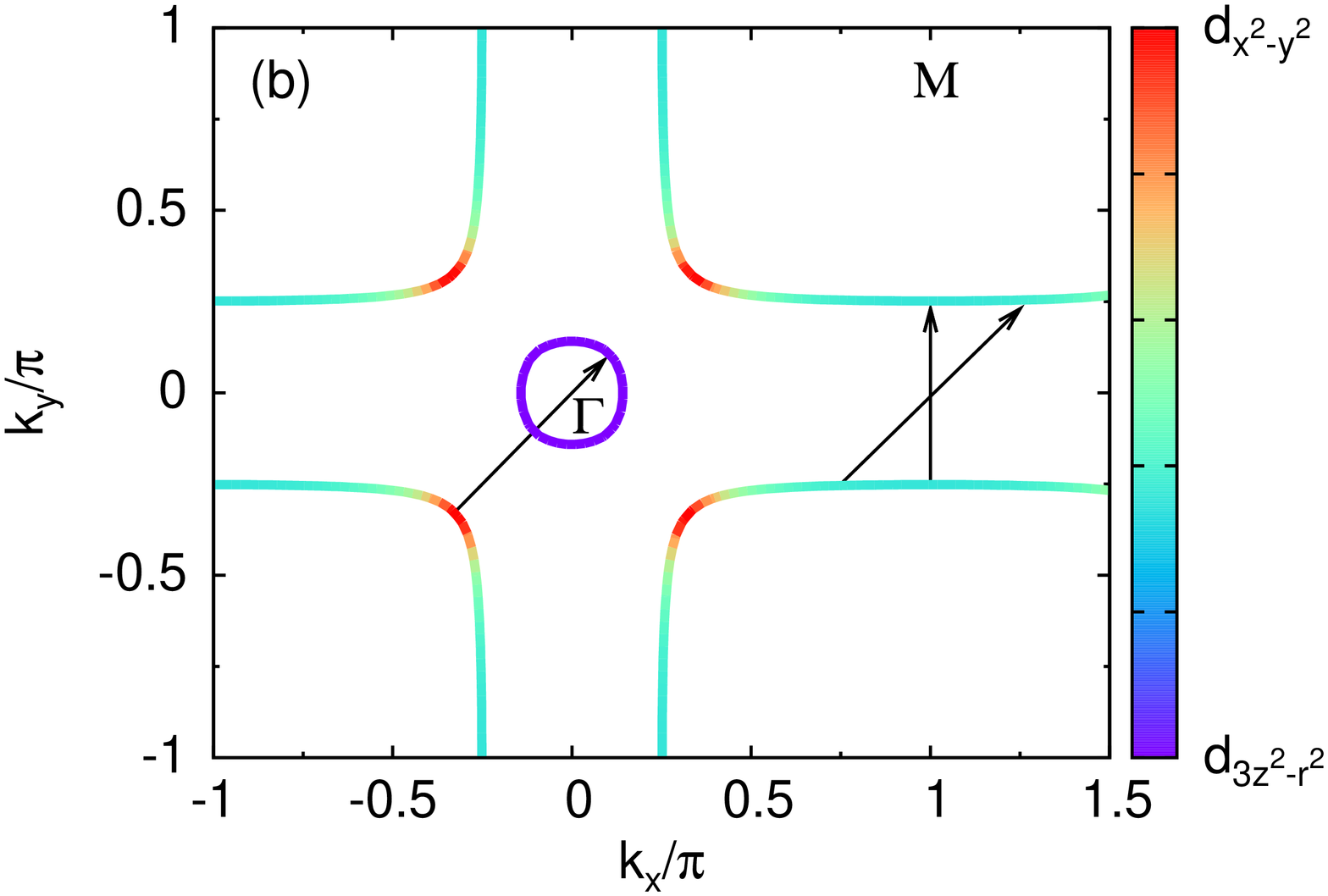,width=55mm,angle=-90}
\vspace*{-5mm}
\end{center}
\caption{Fermi surfaces with orbital densities for different values of the crystal-field parameter $\Delta$ = (a) 0.0 and (b) $-0.3$ with the chemical potentials $\mu = -1.21$ and $-1.245$, respectively.}
\label{fermis}
\end{figure}  
\begin{figure}
\begin{center}
\vspace*{-2mm}
\hspace*{0mm}
\psfig{figure=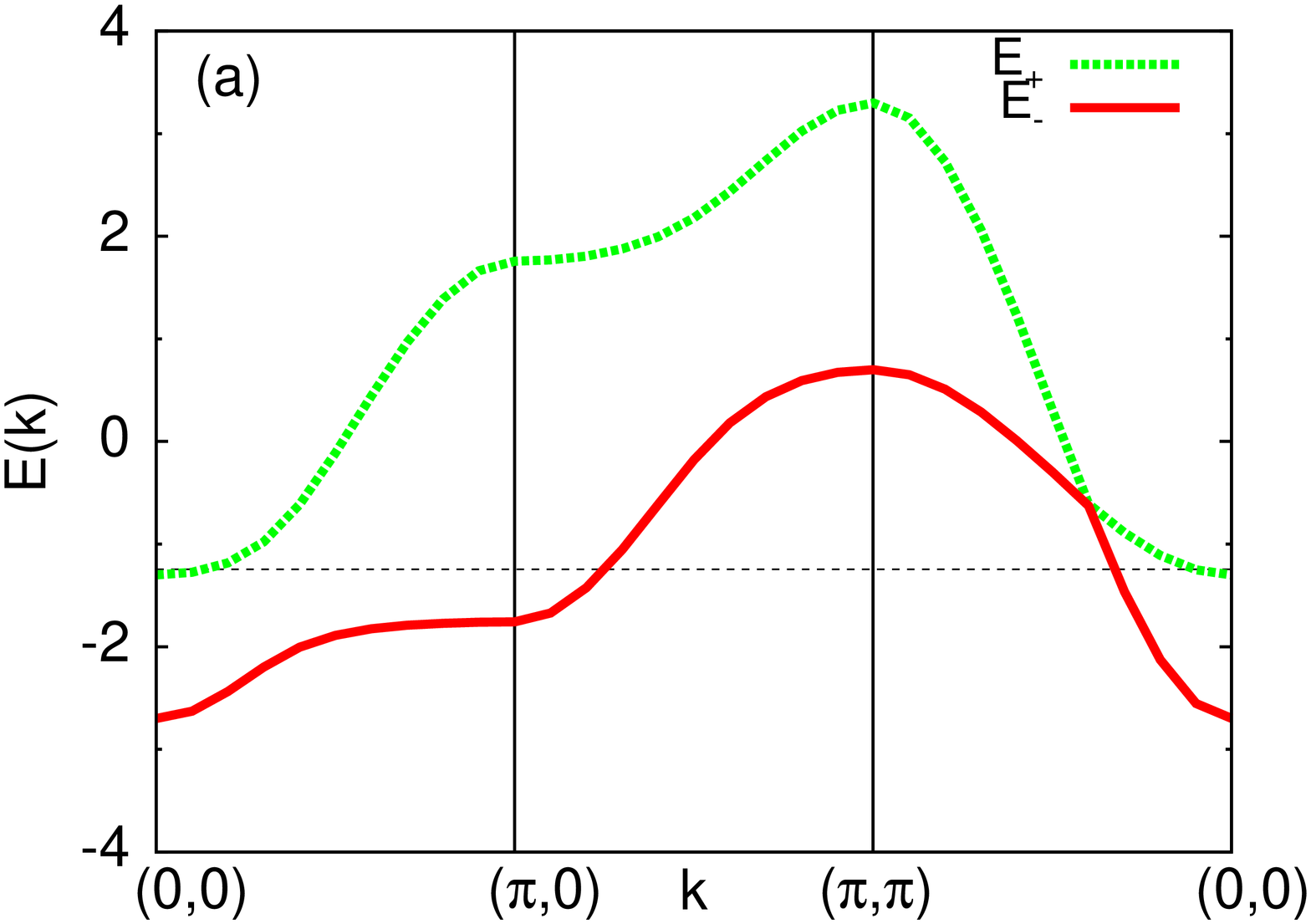,width=50mm,angle=-90}
\hspace*{-8mm}
\psfig{figure=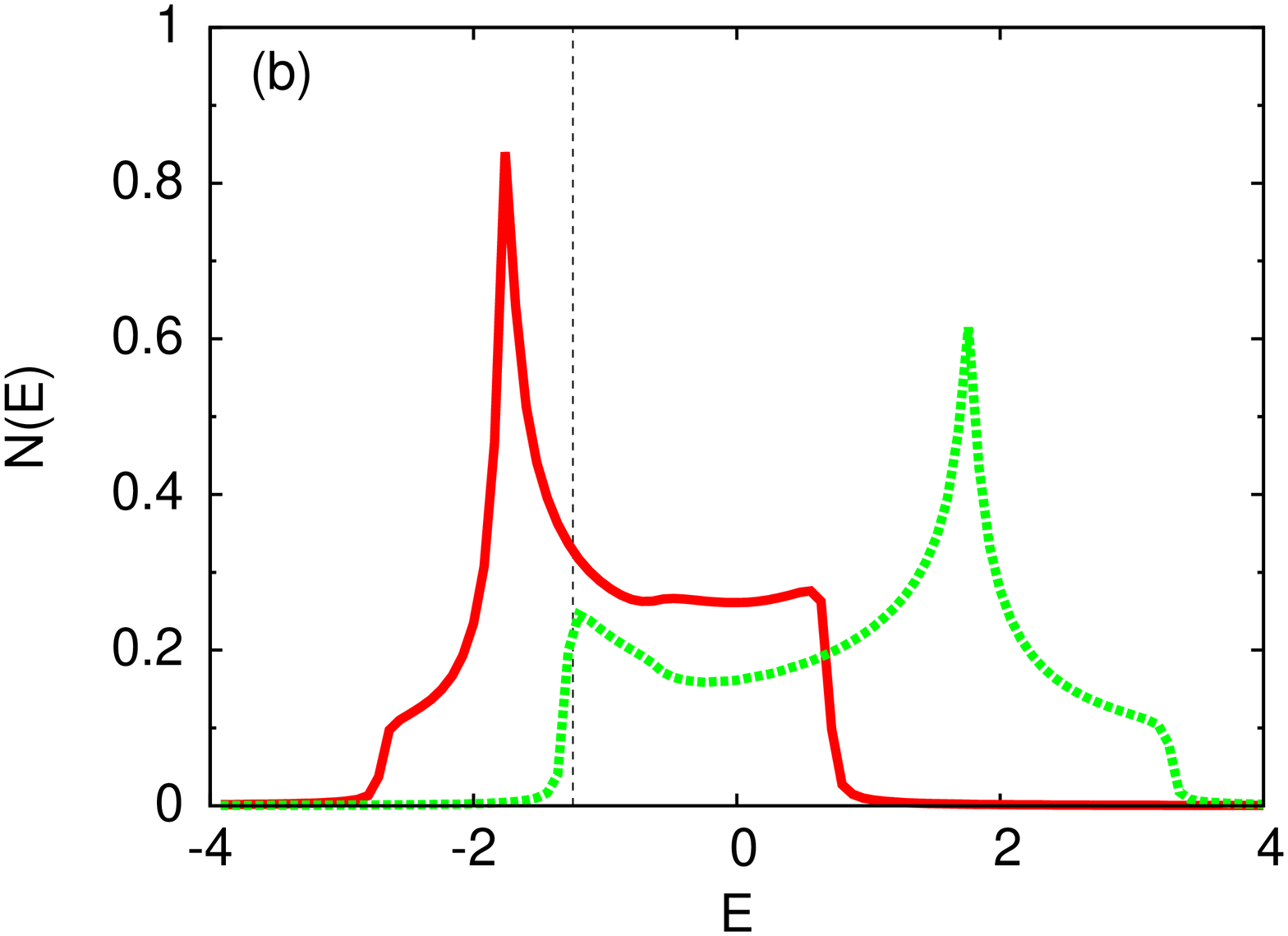,width=55mm,angle=-90}
\vspace*{-5mm}
\end{center}
\caption{(a) Electronic dispersion in the high-symmetry directions with the crystal-field splitting $\Delta = -0.3$ and chemical potential $\mu = -1.245$ , and (b) density
 of states for the two bands with each having a Van Hove singularity.}
\label{dens}
\end{figure}  
The kinetic and the CEF parts of the Hamiltonian can be expressed as
\begin{equation}
H_{\rm kin}({\bf k})+H_{\rm CEF}=\sum_{{\bf k} \sigma}\psi^+_{{\bf k} \sigma}\left[\left(\varepsilon_+({\bf k})-\mu\right) \hat{\tau}^0 +\varepsilon_-({\bf k})
\hat{\tau}^3+\varepsilon_{12}({\bf k})\hat{\tau}^1\right]\psi_{{\bf k} \sigma},
\end{equation}
where $\psi^{\dag}_{{\bf k} \sigma}=(d_{1 \sigma}({\bf k}) \,\, d_{2 \sigma}({\bf k}))$, $\mu$ is the chemical potential, $\hat{\tau}^0$, $\hat{\tau}^1$, and $\hat{\tau}^3$ are the unit matrix, $x$, and $z$ components of the 
Pauli matrices, respectively. The coefficients of the components are given by 
\begin{eqnarray}
  \varepsilon_+({\bf k}) & = & (\varepsilon_1({\bf k})+\varepsilon_2({\bf k}))/2, \,\varepsilon_-({\bf k})  =  (\varepsilon_1({\bf k})-\varepsilon_2({\bf k}))/2-\Delta \nonumber\\
     \varepsilon_{12}({\bf k}) & = & \sqrt{3}(\cos k_x-\cos k_y)/2\nonumber,
\end{eqnarray}
with
\begin{eqnarray}
    \varepsilon_1({\bf k}) & = & -3(\cos k_x+\cos k_y)/2,\,
    \varepsilon_2({\bf k})  =  -(\cos k_x+\cos k_y)/2.\nonumber 
\end{eqnarray}
Then, the matrix of the single electron Matsubara Green's function is described by
\begin{equation}
\hat{G}^{(0)} ({\bf k}, i \omega_n ) = \frac{\left( i \omega_n - \varepsilon_+({\bf k}) +\mu \right) \hat{\tau}^0 - \varepsilon_-({\bf k}) 
\hat{\tau}^3 + \varepsilon_{12} ({\bf k}) \hat{\tau}^1}{\left(i \omega_n - E_+({\bf k}) \right) \left(i \omega_n - E_-({\bf k}) \right)},
\end{equation}
where
\begin{equation}
E_{\pm}({\bf k}) = \varepsilon_+({\bf k}) \pm \sqrt{\varepsilon_-^2({\bf k}) +
\varepsilon_{12}^2({\bf k})} - \mu\label{Ek},
\end{equation}
and Fermionic Matsubara frequency $\omega_n=(2n+1)\pi T$. 

Fig. \ref{fermis} shows the Fermi surfaces for different values of crystal-field parameter $\Delta =  0.0$ and $-0.3$ with chemical potentials $\mu = -1.21$ and $-1.245$, respectively. For both the cases, a large hole pocket is present around the M point. However, a circular electron pocket around the $\Gamma$ point and relatively straightened hole pockets around the M point are the characteristic
 features for $\Delta =  -0.3$, which are in agreement with the ARPES measurement on the half-doped La$_{0.5}$Sr$_{1.5}$MnO$_4$.\cite{evtushinsky} The electron pocket has predominantly 
$d_{3z^2-r^2}$ orbital character while the hole pockets have large $d_{x^2-y^2}$ ($d_{3z^2-r^2}$) orbital
 character in the $\Gamma$-M
 ($\Gamma$-X) direction. Especially, since the orbital mixing ($\varepsilon_{12}({\bf k})$) vanishes along $\Gamma$-M, the electronic states on the electron and hole Fermi surfaces have 
the orbital characters of $d_{3z^2-r^2}$ and $d_{x^2-y^2}$ (Fig. \ref{fermis}), respectively. Two types of nesting vector ($0.5\pi$, 0) and ($0.5\pi$, $0.5\pi)$ are 
expected for $\Delta =  -0.3$ because of the enhanced flatness of the hole Fermi Surface and the additional Fermi surface around the $\Gamma$ point, respectively. The former is
 the nesting between Fermi surface segments dominated by the same orbital $d_{3z^2-r^2}$, while the latter is the one between Fermi surfaces dominated by different orbitals.  
 Fig. \ref{dens}
 shows the electron band-structure and the density of states with $\Delta = -0.3$ and $\mu = -1.245$. There is a Van Hove singularity corresponding to a saddle point at ${\bf k}$ = ($\pi$, 0) for each band.
\section{Static Susceptibility}
To investigate the orbital-ordering instabilities, we consider the orbital susceptibilities defined as follows:
\begin{equation}
\chi^{ij}(\q,i\Omega_n)= \int^{\beta}_0{d\zeta e^{i \Omega_{n}\zeta}\langle T_\zeta [{\cal T}^i_\q(\zeta) {\cal T}^j_{-\q}(0)]\rangle}.
\end{equation}
Here, $\langle...\rangle$ denotes thermal average, $T_\zeta$ imaginary time ordering, and $\Omega_n$ are the Bosonic
Matsubara frequencies. 
${\cal T}^l_{\bf q}$ is 
obtained as the Fourier transformation of 
$\mathcal{T}^l_{{\bf i}}$ described in the previous section 
\begin{equation}
{\cal T}^l_{\bf
q}=\sum_{\sigma {\bf k}}\psi_{\sigma }^\dagger({\bf
k+q})\hat{\tau}^l\psi_{\sigma}({\bf k}),
\end{equation}
where, $\psi_{\sigma}^{\dag}({\bf k})=(d^{\dagger}_{1 \sigma}({\bf k}), \,\, d^{\dagger}_{2 \sigma}({\bf k}))$.
For the spin-ordering instability, 
it will be 
sufficient to replace the orbital operators in Eq. (12) by the $z$-component of the spin operator defined as
\begin{equation}
 {S}^{zl}_{\bf
q}=\frac{1}{2}\sum_{ \sigma {\bf k}}\psi_{ \sigma }^\dagger({\bf
k+q})\sigma \hat{\tau}^l_{}\psi_{ \sigma}({\bf k}),
\end{equation}
where $\sigma$ = $\pm1$.

Divergence of the static spin (orbital) susceptibility ($\hat{\chi}^{\rm s(o)}({\bf q})$) enhanced by the Coulomb interaction $\hat{U}^{\rm s(o)}$ signals
 the spin-ordering (orbital-ordering) instability,\cite{takimoto} which is calculated within the RPA-level
\begin{eqnarray}
    && \hat{\chi}^{\rm s(o)}({\bf q}) \!=\!
  [\hat{1}+\hat{U}^{\rm s(o)}\hat{\chi}({\bf q})]^{-1}
  \hat{\chi}({\bf q}),
\end{eqnarray}
where row and column labels appear in the order
11, 22, 12, and 21 with 1 and 2 being the orbital indices. $\hat{1}$ is the 4$\times$4 unit matrix.
The matrix elements of $\hat{\chi}({\bf q})$ are defined by
$\chi_{\mu\nu,\alpha\beta}({\bf q})$=
$-T\sum_{{\bf k},n}
G^{(0)}_{\alpha\mu}({\bf k}+{\bf q},i\omega_{n})
G^{(0)}_{\nu\beta}({\bf k},i\omega_{n})$, 
$G^{(0)}_{\mu\nu}({\bf k},i\omega_{n})$
is the bare Green's function 
given in the last section. 
After integrating out the Jahn-Teller phonons, the Coulomb interaction
 $\hat{U}^{\rm s(o)}$ is replaced by the renormalized one $\hat{\tilde{U}}^{s(o)}$ given by \cite{kontani} 
\begin{equation}
\tilde{U}^{s(o)}_{n_1 n_2, n_3 n_4}= \left\{
\begin{array}{@{\,} l @{\,} c}
-U~(U-2g^{\prime 2}_0 D_0(i\Omega_n)-2g^{\prime 2}_z D_z(i\Omega_n)) & (n_1=n_2=n_3=n_4)\\
-U'~(-U'- 2g^{\prime 2}_x D_x(i\Omega_n)) & (n_1=n_3\ne n_2=n_4)\\
0~(2U'-2g^{\prime 2}_0 D_0(i\Omega_n)+2g^{\prime 2}_z D_z(i\Omega_n)) & (n_1=n_2\ne n_3=n_4)\\
0~(-2g^{\prime 2}_x D_x(i\Omega_n))& (n_1=n_4\ne n_2=n_3)\\
0 & (\mathrm{otherwise})
\end{array}\right.,
\end{equation}
where the local phonon Green's function is
\begin{equation}
 D_l(i\Omega_n)=\frac{2\omega^{\prime}_l}{\Omega^{2}_{n}+\omega^{\prime 2}_l}.
\end{equation}
In the following, we use a dimensionless electron-phonon coupling as $\lambda = 2g^{\prime 2}_x D_x(0)$.

The instability for the spin- and orbital-ordered phases
is determined by the conditions ${\rm det}[\hat{1}+\hat{\tilde{U}}^{\rm s(o)}\hat{\chi}({\bf q})]$ = 0. Especially, the instability equation for $q_x = q_y$ reduces to ${\rm det}[\hat{1}+\hat{\tilde{U}}^{\rm s(o)}_1\hat{\chi}_1({\bf q})]$ $\times$
${\rm det}[\hat{1}+\hat{\tilde{U}}^{\rm s(o)}_2\hat{\chi}_2({\bf q})]$ = 0 due to the block-diagonal forms of both the interaction $\hat{\tilde{U}}^{s(o)}$ and the bare susceptibility 
$\hat{\chi}({\bf q})$ matrices, where $\hat{\tilde{U}}^{\rm s(o)}_i$ and $\hat{\chi}_i({\bf q})$ are 2 $\times$ 2 matrices in the basis spanned by 11 and 22 (12 and 21) for $i = 1\vspace{2mm}(2)$. The momentum dependence
 of all the matrix elements of $\hat{\chi}_i({\bf q})$ are of A$_{1g}$ representation, while those of remaining matrix elements of $\hat{\chi}({\bf q})$ are of the B$_{1g}$ representation, whose
 basis function is $q^2_x - q^2_y$. Therefore, the elements of the off-diagonal 2$\times$2 block matrices vanish identically for $q_x = q_y$.

In the tetragonal symmetry, orbital operators ${\cal T}^0$, ${\cal T}^x$, ${\cal T}^y$, and ${\cal T}^z$, belong to the one dimensional representations A$_{1g}$, B$_{1g}$, A$_{2g}$, and A$_{1g}$, respectively. Here,
 we note that ${\cal T}^x$ breaks the four-fold rotation symmetry whereas ${\cal T}^z$ does not. The transition to the ordered state with A$_{1g}$ symmetry formed by the
 two order parameters  ${\cal T}^0_{\bf q}$ and ${\cal T}^z_{\bf q}$ is signaled by the divergence of charge susceptibility $\chi^{00}(\q)$ (charge 
instability) or longitudinal orbital susceptibility $\chi^{zz}(\q)$ (orbital instability of the type $d_{x^2-y^2}$/$d_{3z^2-r^2}$) corresponding to
 the condition ${\rm det}[\hat{1}+\hat{\tilde{U}}^{\rm o}_1\hat{\chi}_1({\bf q})]$ = 0. On the other hand, the transition to
 the orbital ordered state with the order parameter ${\cal T}^x_{\bf q}$ belonging to the  B$_{1g}$ symmetry is signaled by the 
divergence of transverse orbital susceptibility $\chi^{xx}(\q)$ corresponding to the condition ${\rm det}[\hat{1}+\hat{\tilde{U}}^{\rm o}_2\hat{\chi}_2({\bf q})]$ = 0. 
To see the orbital ordering type for the transversal one, 
fermion operators are transformed as $d_{\pm}=\frac{1}{\sqrt{2}}(-d_{3z^2-r^2}\,{\pm}\,d_{x^2-y^2}$), which follows essentially from a rotation of $\theta = \pi/2$ in the orbital space given by
  $\left|{\pm}\right\rangle = - \cos (\theta/2)\left|{3z^2-r^2}\right\rangle \pm \sin (\theta/2)\left|{x^2-y^2}\right\rangle $. In the new orbital basis
  $\left|+\right\rangle$ and 
 $\left|-\right\rangle$, operator ${\cal T}^x_{\bf q}$ can be expressed as 
\begin{equation}
{\cal T}^x_{\bf
q}=-\sum_{ \sigma  {\bf k}}\Psi_{  \sigma}^\dagger({\bf
k+q})\hat{\tau}^{z^{\prime}}_{}\Psi_{ \sigma }({\bf k}),
\end{equation}
with $\Psi_{\sigma}^{\dag}({\bf k})=(d^{\dagger}_{+ \sigma}({\bf k}), \,\, d^{\dagger}_{- \sigma}({\bf k}))$. ${\cal T}^y$ has not been considered here because it breaks time-reversal symmetry and is not associated with any orbital moment.

\begin{figure}
\begin{center}
\vspace*{-2mm}
\hspace*{0mm}
\psfig{figure=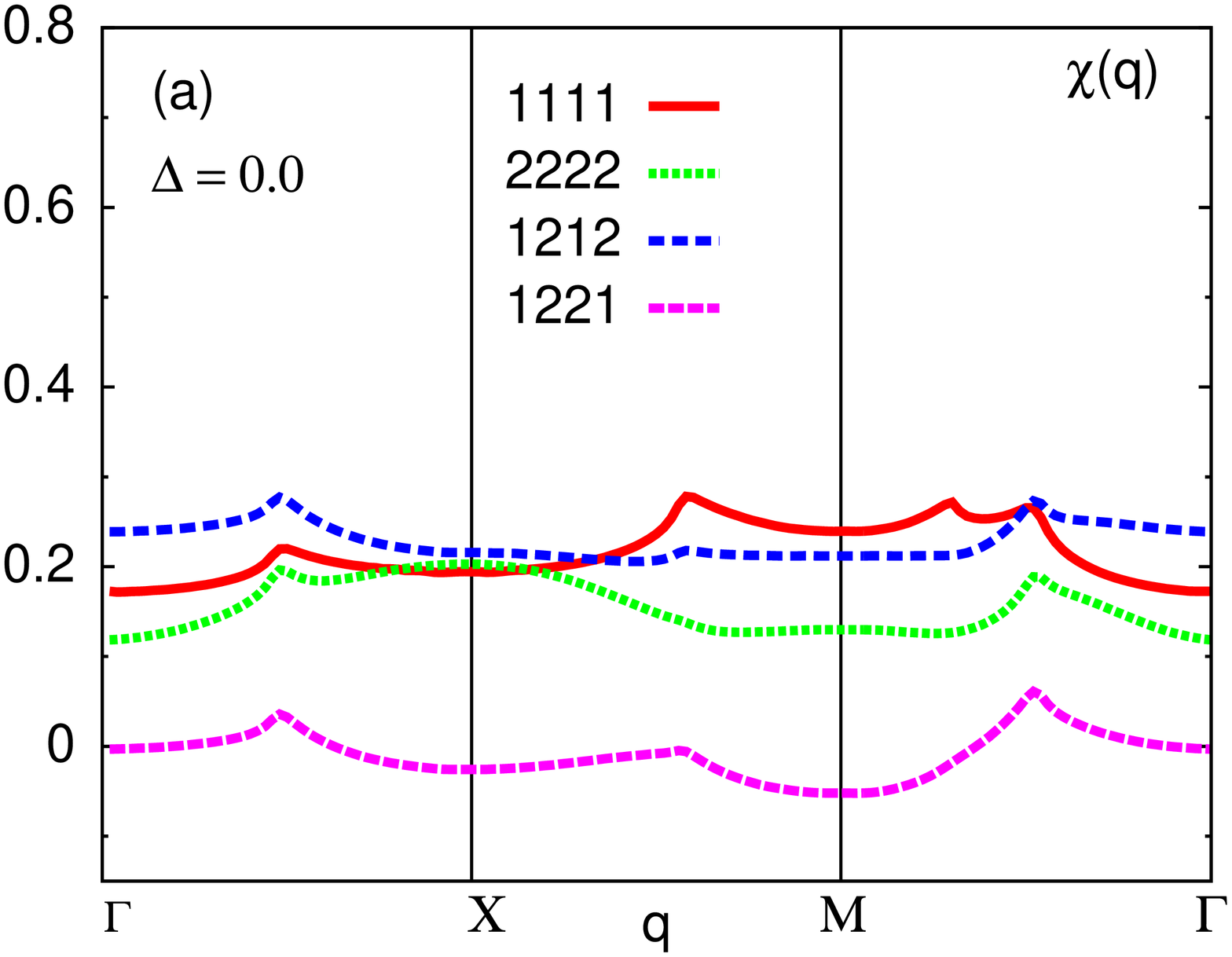,width=55mm,angle=-90}
\psfig{figure=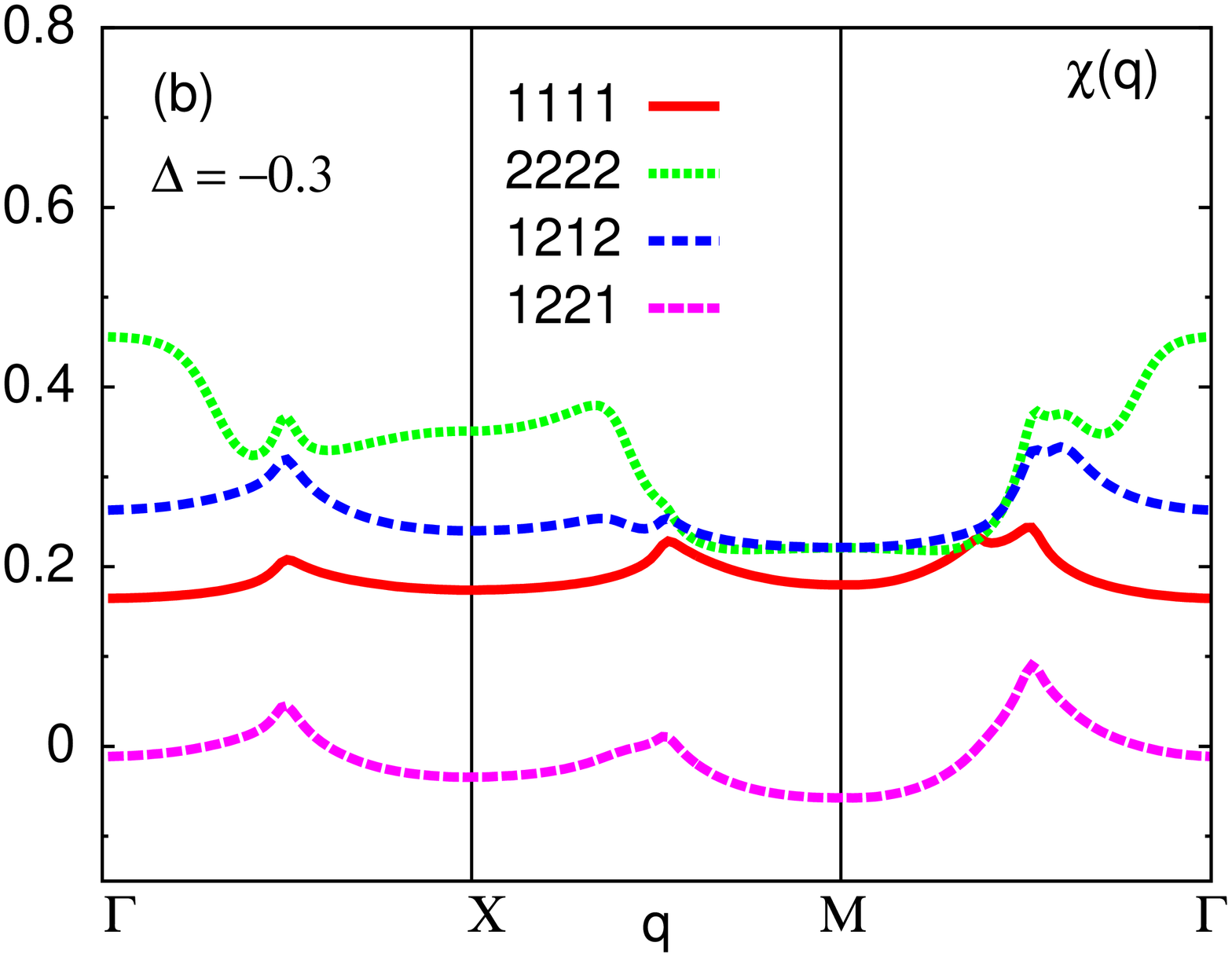,width=55mm,angle=-90}
\vspace*{-5mm}
\end{center}
\caption{Principal components of one loop susceptibility for $\Delta$ = (a) 0.0 and (b) $-0.3$.}
\label{sus}
\end{figure}  
Fig. \ref{sus} (a) and (b) show the key components of 
one bubble susceptibility 
$\hat{\chi}({\bf q})$ for $\Delta = 0.0$ and $-0.3$, respectively. For $\Delta$ = 0.0, the components exhibit very weak peak 
structures at $(0.5 \pi, 0)$ and $(0.5 \pi, 0.5 \pi)$. On the other hand, three features emerge for $\Delta = -0.3$, 
the peaks of the components $\chi_{1212}({\bf q})$ and $\chi_{1221}({\bf q})$ contributing to the transverse orbital susceptibility are enhanced at $(0.5 \pi, 0)$ and $(0.5 \pi, 0.5 \pi)$ because of the improved nesting between different Fermi surfaces, 
and two additional peaks appear for $\chi_{2222}({\bf q})$ at $\approx (\pi$, $0.3\pi$) and (0, 0) due to the small electron pocket. Therefore, depending on the relative 
strengths of the intra- and inter-orbital renormalized Coulomb interactions, the intra- and inter-pocket scatterings will lead either to the ferromagnetic 
 instability or to the orbital instability.
\begin{figure}
\begin{center}
\vspace*{-2mm}
\hspace*{0mm}
\psfig{figure=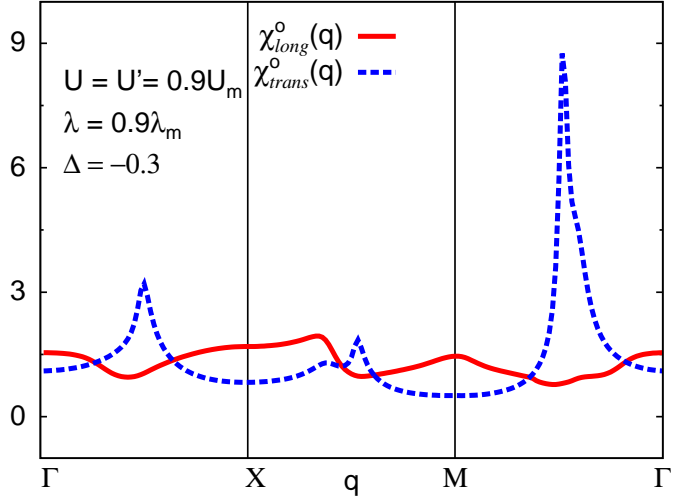,width=70mm,angle=-90}
\vspace*{-5mm}
\end{center}
\caption{RPA-level orbital susceptibilities for $U = U'= 0.9U_m$, $\lambda = 0.9\lambda_m$, and $f = 0.5$, where $\Delta = -0.3$. }
\label{rpa}
\end{figure}  

 Fig. \ref{rpa} shows orbital susceptibilities within RPA calculated at $U = U' = 0.9 U_m = 1.35$, $\lambda = 0.9 \lambda_m = 0.405$, $f = 0.5$,\cite{hotta} and $\Delta = -0.3$, where interaction
 parameters with the subscript $m$ denote the critical strength 
at which susceptibilities diverge. The transverse-orbital susceptibility is enhanced due to the electron correlations at $(0.5 \pi, 0.5 \pi)$, thereby implying instability for the transversal orbital 
ordering of $d_+$/$d_-$ type.
 \section{Phase Diagram}
\begin{figure}
\begin{center}
\vspace*{-2mm}
\hspace*{0mm}
\psfig{figure=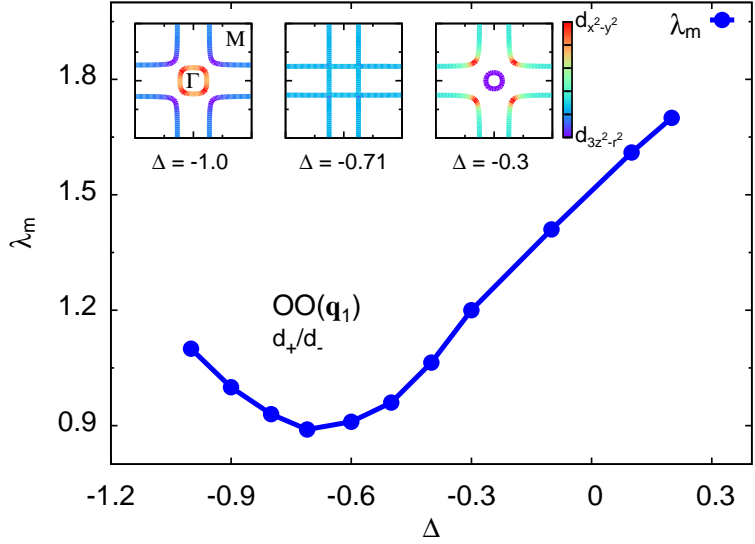,width=100mm,angle=-0}
\vspace*{-5mm}
\end{center}
\caption{Dependence of the critical value of electron-phonon coupling $\lambda_m$ on the crystal field parameter $\Delta$ with $U = U' = 0$. OO(${\bf q}_1$) is the orbitally ordered state with wavevector ${\bf q}_1$
 = ($0.5 \pi, 0.5 \pi$). The inset 
shows the Fermi surfaces with orbital densities for different values of $\Delta$ = $-1.0$, $-0.71$, and $-0.3$. }
\label{crys}
\end{figure}  
To discuss the role of crystal-field splitting $\Delta$ on the orbital ordering, we consider a simplified model with only electron-phonon coupling involving Jahn-Teller distortions 
of ($x^2-y^2$)- and ($3z^2-r^2$)-type $(U = U' = 0)$. Fig. \ref{crys} shows the critical value of electron-phonon coupling $\lambda_m$ as a function of
 $\Delta$, which causes a transition to the $d_{x^2-z^2}$/$d_{y^2-z^2}$-type orbitally ordered state OO(${\bf q}_1$) with wavevector ${\bf q}_1$ = ($0.5\pi$, $0.5\pi$). $\lambda_m$ has a minimum at
$\Delta = -1/\sqrt{2}$, resulting from the optimal nesting which in turn follows from the flat segments of the hole pockets. This can be seen by substituting, for instance, $k_x = \pi/4$ in the 
electron dispersion (Eq. (11)), which yields $E_{\pm}({\bf k}) = -\sqrt{2}$ independent of $k_y$ for $\Delta = -1/\sqrt{2}$. In addition to the weakened nesting, the factor which is also responsible for the increase of
 critical $\lambda_m$ away from $\Delta = -1/\sqrt{2}$
 is the increased proportion of $d_{3z^2-r^2}$ ($d_{x^2-y^2}$) orbital in the hole pocket for $\Delta < -1/\sqrt{2}$ ($\Delta > -1/\sqrt{2}$), and vice-versa in the case of electron pocket.
\begin{figure}
\begin{center}
\vspace*{-2mm}
\hspace*{0mm}
\psfig{figure=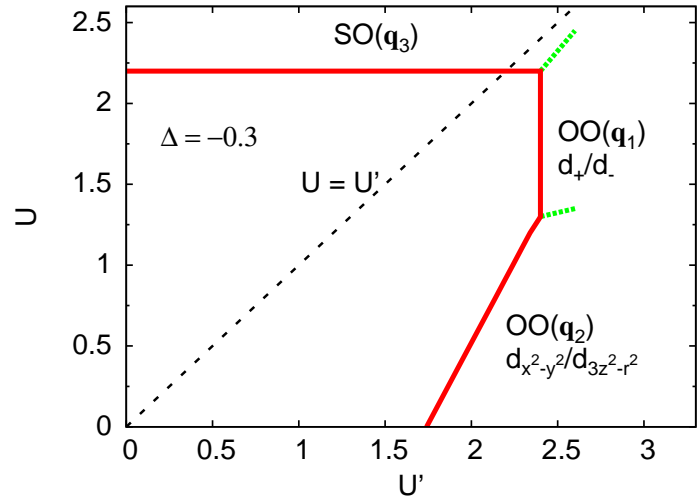,width=70mm,angle=-90}
\vspace*{-5mm}
\end{center}
\caption{$U-U'$ phase diagram. SO and OO are abbreviations for the spin ordered and orbitally ordered, respectively. ${\bf q}_1$ = ($0.5 \pi, 0.5 \pi$), ${\bf q}_2 \approx (\pi, 0.25\pi) $, and ${\bf q}_3$ = (0, 0).}
\label{phasda}
\end{figure}

Fig. \ref{phasda} shows the $U-U'$ phase diagram obtained from the instability analysis for $\lambda = 0$, which consists of states with the ordering wavevector mentioned in the previous section. There are three 
ordered regions, transversal $d_{x^2-z^2}$/$d_{y^2-z^2}$-type orbitally
 ordered state OO(${\bf q}_1$) with wavevector ${\bf q}_1$ = ($0.5\pi$, $0.5\pi$), longitudinal $d_{x^2-y^2}$/$d_{3z^2-r^2}$-type orbitally ordered state OO(${\bf q}_2$) with wavevector ${\bf q}_2 \approx (\pi$, $0.3\pi$), and ferromagnetic
 state SO(${\bf q}_1$) with ${\bf q}_3$ = ${\bf 0}$. The instabilities are caused by the peaks of $\chi_{2222}({\bf q})$,
 $\chi_{1212}({\bf q})$ and $\chi_{1221}({\bf q})$, and $\chi_{2222}({\bf q})$ at ${\bf q}$ = ${\bf q}_1$, ${\bf q}_2$, and ${\bf q}_3$, respectively. The straight phase-boundary lines of OO(${\bf q}_1$)
 and SO(${\bf q}_3$) states, corresponding to the
 largest eigenvalues of -$\hat{U}^{\rm o}_2\hat{\chi}_2({\bf q}_1)$ and -$\hat{U}^{\rm s}_1\hat{\chi}_1({\bf q}_3)$ being unity, result from
 the diagonal interaction matrices $\hat{U}^{\rm o}_2$ and $\hat{U}^{\rm s}_3$ with equal elements, respectively. The critical line for the OO(${\bf q}_2$) corresponds to the largest eigenvalue 
of -$\hat{U}^{\rm o}\hat{\chi}({\bf q}_2)$ being unity for $q_x \neq q_y$. 
It should be noted that only SO(${\bf q}_3$) is obtained in the realistic 
parameter region $U \ge U'$. 

\begin{figure}
\begin{center}
\vspace*{-2mm}
\hspace*{0mm}
\psfig{figure=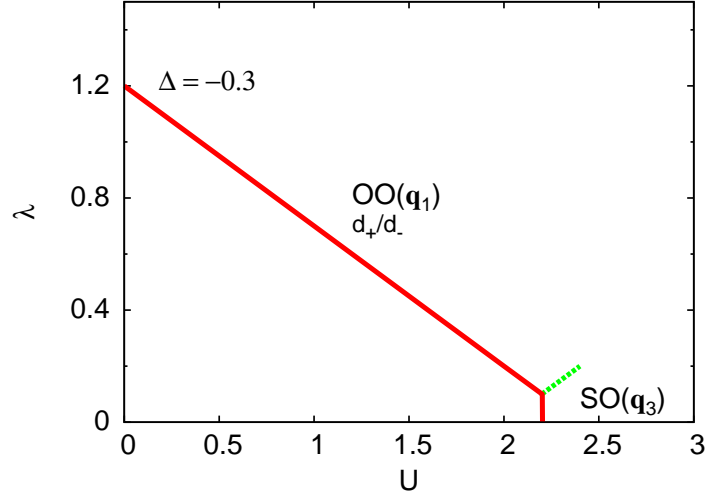,width=70mm,angle=-90}
\vspace*{-5mm}
\end{center}
\caption{$\lambda-U$ phase diagram with $U=U'$.}
\label{phasdb}
\end{figure}

Fig. \ref{phasdb} shows the $\lambda-U$ phase diagram, where $U=U'$ for simplicity as the on-site Coulomb interactions have the similar order of magnitudes. According 
to the $U-U'$ phase diagram, the SO(${\bf q}_3$) phase appears on increasing $U$ along $U = U'$. Since $\hat{\tilde{U}}^{s}$ is independent of the Jahn-Teller coupling,
 a vertical phase-boundary line for the SO(${\bf q}_3$) is obtained 
in the $\lambda-U$ phase diagram. 
However, OO(${\bf q}_1$) is readily stabilized than the SO(${\bf q}_3$) state by the Jahn-Teller coupling. The critical straight
 line for the OO(${\bf q}_1$) arises due to the non-diagonal symmetric interaction matrix $\hat{\tilde{U}}^{\rm o}_2$ with equal diagonal elements, so that the matrix product
 of the interaction and the bare susceptibility is also symmetric with equal diagonal elements, which yields a linear relation between $\lambda_m$ and $U_m$.

The inclusion of the intra-orbital Coulomb interaction $U$ is essential 
to realize OO(${\bf q}_1$) (Fig. \ref{phasda}). Importantly, as $U$ is always larger than $U'$, it is impossible to obtain OO(${\bf q}_1$) by considering 
only the electronic correlations in the appropriate parameter regime. On the other hand, Jahn-Tellar distortion can stabilize the OO(${\bf q}_1$)
in a large region of interaction parameter space (Fig. \ref{phasdb}). However, the region of
 stability may get reduced by the double-exchange mechanism
 which supports delocalization induced ferromagnetism.\cite{zener}
  
\begin{figure}
\begin{center}
\vspace*{-2mm}
\hspace*{0mm}
\psfig{figure=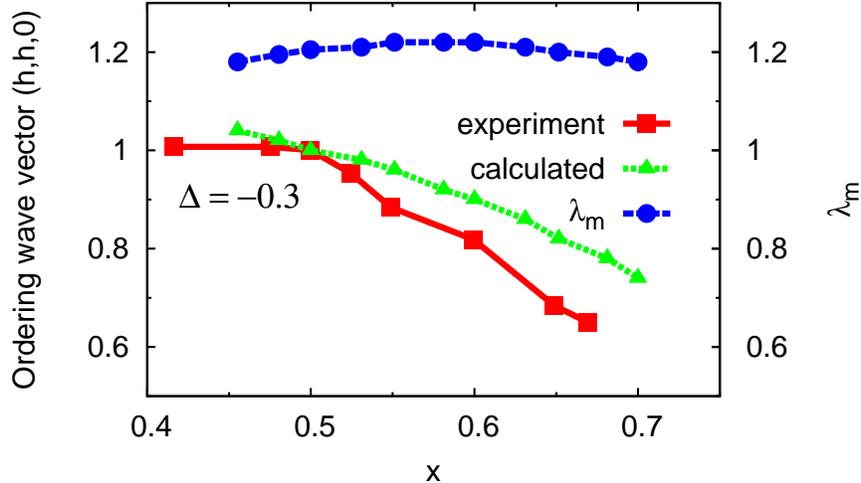,width=70mm,angle=-90}
\vspace*{-5mm}
\end{center}
\caption{Comparison of the doping dependence of calculated orbital-ordering wavevector by using a simplified model ($U = U' =0$)
with the experimental data obtained from x-ray scattering experiments.\cite{Larochelle} Corresponding value of the critical electron-phonon coupling $\lambda_m$ is also shown. }
\label{dop}
\end{figure}  
\vspace*{1\baselineskip}
\section{Conclusions and Discussions}
Our investigation has highlighted the roles of crystal-field splitting and Jahn-Teller distortions in
the orbital ordering transition to the charge-orbital ordered state near half doping in the single-layer manganites. 

Crystal-field splitting of energy levels favoring $d_{3z^2-r^2}$ orbital occupancies over $d_{x^2-y^2}$ reproduces the Fermi
 surface of La$_{0.5}$Sr$_{1.5}$MnO$_4$ as observed in the ARPES experiments, having a circular electron pocket mainly composed of $d_{3z^2-r^2}$ orbital 
around the $\Gamma$ point within a tight-binding model including only nearest-neighbor hopping. Moreover, the Fermi surface nesting is strengthened
 due to the flattening of hole pockets around the M point, which has predominantly $d_{x^2-y^2}$ ($d_{3z^2-r^2}$) orbital 
character in the $\Gamma$-M ($\Gamma$-X) direction. 

Orbitally ordered state with ordering wavevector ($0.5\pi$, $0.5\pi$) observed in La$_{0.5}$Sr$_{1.5}$MnO$_4$ cannot be reproduced in a purely
 electronic model with the intra- and inter-orbital Coulomb interactions only unless the former is smaller than the latter. However, the unphysical situation can be avoided by including Jahn-Teller distortions to realize 
the orbital-ordering transition 
of the transversal $d_+/d_-$-type breaking the four-fold rotation symmetry with character similar to the ordering observed in several experiments using resonant elastic soft x-ray scattering and linear dichroism. The breaking of the four-fold rotation symmetry, on the other hand, can be detected by carrying out the torque measurements.\cite{kasahara} 

Finally, we discuss the doping dependence of the orbital ordering wavevector observed by the x-ray measurements on La$_{0.5}$Sr$_{1.5}$MnO$_4$. Fig. \ref{dop} shows a comparison of the calculated orbital 
ordering wavevector as a function of hole doping with the experimental data for La$_{1-x}$Sr$_{1+x}$MnO$_4$
obtained from x-ray experiments.\cite{Larochelle} The linear dependence of the ordering wavevector in the region 0.5 $\leq$ $x$ $\leq$ 0.7 is in agreement with the experiment except for a slight deviation in the slope. However, 
the agreement is poor for $x < 0.5$ where the experimental ordering wavevector is almost constant due to a possible phase separation.\cite{Larochelle} The critical 
value of electron-phonon coupling $\lambda_m$ shows a little variation as a function of carrier concentration, while the staggered orbital ordering is always of the type $d_+/d_-$ as in the case of half doping. A simultaneous 
charge-orbital ordering may be obtained by including a long-range Coulomb interaction appropriate for the charge-density wave with the wavevector 2${\bf Q}$, where ${\bf Q}$ is the orbital ordering wavevector.

\section* {Acknowledgements}

The work of T. Takimoto is supported by Basic Science Program through the National Research Foundation of Korea (NRF) funded by the Ministry of Education (NRF-2012R1A1A2008559). 
D. K. Singh would like to acknowledge the Korea Ministry of Education, Science and Technology, Gyeongsangbuk-Do and Pohang City for the support of the Young Scientist Training program at the
Asia-Pacific Center for Theoretical Physics. The authors thank K. H. Lee for useful discussions.

\end{document}